\documentclass[manuscript]{aastex}

\shorttitle{The pulsating, eclipsing binary OO Dra}

\shortauthors{Zhang et al.}

\begin{document}

\title{Photometric study of the pulsating, eclipsing binary OO Dra}

\author{X.B. Zhang\altaffilmark{1}, L.C. Deng\altaffilmark{1}, J.F. Tian\altaffilmark{1}, K. Wang\altaffilmark{1}, J.J. Sun\altaffilmark{2}, Q.L. Liu\altaffilmark{2}, H.Q. Xin\altaffilmark{2}, Q. Zhou\altaffilmark{2}, Z.Z. Yan\altaffilmark{1,3}, Z.Q. Luo\altaffilmark{3}, C.Q. Luo\altaffilmark{1}}

\affil {1, Key Laboratory of Optical Astronomy, National Astronomical
Observatories, Chinese Academy of Sciences, Beijing, 100012, China}

\affil {2, Purple Mountain Observatory, Chinese Academy of Sciences, Nanjing, 210008, China}

\affil {3, Department of Physics, China West Normal University, Nanchong 637002, China}

\begin{abstract}

We present a comprehensive photometric study of the pulsating, eclipsing binary OO Dra. Simultaneous B- and V-band photometry of the star was carried out on 14 nights. Revised orbital period and a new ephemeris were derived from the data. The first photometric solution of the binary system and the physical parameters of the component stars are determined. It reveals that OO Dra could be a detached system with the less-massive secondary component nearly filling in its Roche lobe.  By subtracting the eclipsing light changes from the data, we obtained the intrinsic pulsating light curves of the hotter and massive primary component. Frequency analysis of the residuals light yields two confident pulsation modes in both B- and V-band data with the dominant frequency detected at 41.865 c/d. A brief discussion concerning the evolutionary status and the pulsation nature of the binary system is finally given.

\end{abstract}

\section{Introduction}

Eclipsing binaries with pulsating components  are important objects to understand stellar structure and evolution. Investigating the frequencies of oscillations, known as asteroseismology, helps us to identify the physical processes behind the pulsating nature and so the stellar interiors. On the other hand, modeling the light and radial velocity curves of a eclipsing binary can precisely determine the physical parameters of the components. This enables one to definitively identify the pulsation modes and compare the results with theoretical models in detail. The study of pulsating stars in eclipsing binaries offers new and strict constraints for stellar theories, helps people to understand the influences of tidal forces and mass transfer among the interacting binaries.

The present study concentrates on the newly discovered pulsating eclipsing binary system OO Dra. The star was first found to be an eclipsing binary with an orbital period of 1.23837 days by Biyalieva \& Khruslov (2007). The pulsation nature of the star was later discovered by Dimitrov et al. (2008) through follow-up observations. With the out-of-eclipse photometric data, they had  made a preliminary frequency analysis and detected a main periodicity about 37 c/d.  A linear ephemeris of the binary system was given as $HJD(MInI)=2451403.832(4)+1.2383832(8) \times E$. In addition, Dimitrov et al. (2008) had contributed a spectroscopy of the binary system. Twelve radial-velocity measurements were given. The spectral type of the primary star in the binary system was identified to be A3 V-IV, with an effective temperature of 8500 K and $logg=4.0$.

As a contribution to the on-going program of searching for and studying of pulsating eclipsing binaries (Zhang et al. 2009, 2013, Liu et al. 2012), we have performed a new photometry of OO Dra and collected sufficient photometric data. We present the results of the observations in this paper, as well as a comprehensive study of the binary system and its intrinsic pulsation.

\section{Photometric observations}

The eclipsing binary OO Dra was observed over 3 weeks in March-April 2014. Time-series CCD photometry were carried out with the 50cm binocular telescope (Deng et al. 2013, Zhang et al. 2014) at the Qinghai Station of the Purple Mountain Observatory, Chinese Academy of Sciences. Data were collected with two Andor 2k$\times$2k CCD, one with a standard Johnson V and the other with a B filter equipped, so that a simultaneous two-color photometry is achieved. Each camera provides a field of view of 20'$\times$20', with an image scale of about 0".59/pixel. Useful data were obtained in 90 hours during 14 nights from  March 19 to April 12, 2014.  A total of 1522 frames in B and 3428 in V filter were collected. The journal of observations is presented in Table 1. 

\begin{table}
\begin{center}
\caption{Observation log.}
\begin{tabular}{lcccc}
\tableline\tableline
Date & Start &  length & Frames (B)  & Frames (V)\\
{} & (HJD2456700+) & (hrs) & {} &{}\\
\tableline
2014 March 22 & 39.134 & 3.8 & 79 & 154 \\
2014 March 23 & 40.049 & 7.8 & 139 & 386 \\
2014 March 24 & 41.042 & 8.0 & 141 & 275 \\
2014 March 25 & 42.024 & 8.1 & 144 & 256 \\
2014 March 27 & 44.044 & 7.8 & 137 & 312 \\
2014 March 31 & 48.069 & 6.5 & 98 & 311 \\
2014 April 01 & 49.055 & 6.8 & 120 & 235 \\
2014 April 02 & 50.095 & 5.9 & 104 & 204 \\
2014 April 03 & 51.048 & 6.9 & 124 & 277 \\
2014 April 04 & 52.054 & 6.7 & 68 & 138 \\
2014 April 05 & 53.077 & 6.1 & 98 & 225 \\
2014 April 06 & 54.104 & 5.4 & 95 & 186 \\
2014 April 07 & 55.030 & 3.4 & 60 & 117\\
2014 April 12 & 60.057 & 6.4 & 115 & 352\\
\tableline
\end{tabular}
\end{center}
\end{table}

The preliminary processing (bias, dark subtraction and flat fielding correction) of the CCD frames was performed with the standard routines of CCDPROC in the IRAF package. Photometry was extracted by using the DAOPHOT II package. Following Dimitrov et al. (2008), we used the stars BD+75 453 (=SAO 7402) and BD+75 415 (=GSC4050-1520) as the comparison and check stars. The magnitude difference between the two stars was confirmed to be stable within 0.015 mag during the observation. The differential magnitudes of OO Dra were then extracted. The time of each measurement was converted to Heliocentric Julian date. As an example, we plot in the upper panels of Figure 1 a part of the time-series light curves . The left one represents the B- and V-band measurements obtained on March 24. Those shown in the right one were derived on April 6. Where the short-term pulsational light variations in addition to the eclipsing light changes can be clearly seen. It confirms the discovery of Dimitrov et al. (2008).
\begin{figure}
\includegraphics[angle=270,scale=.6]{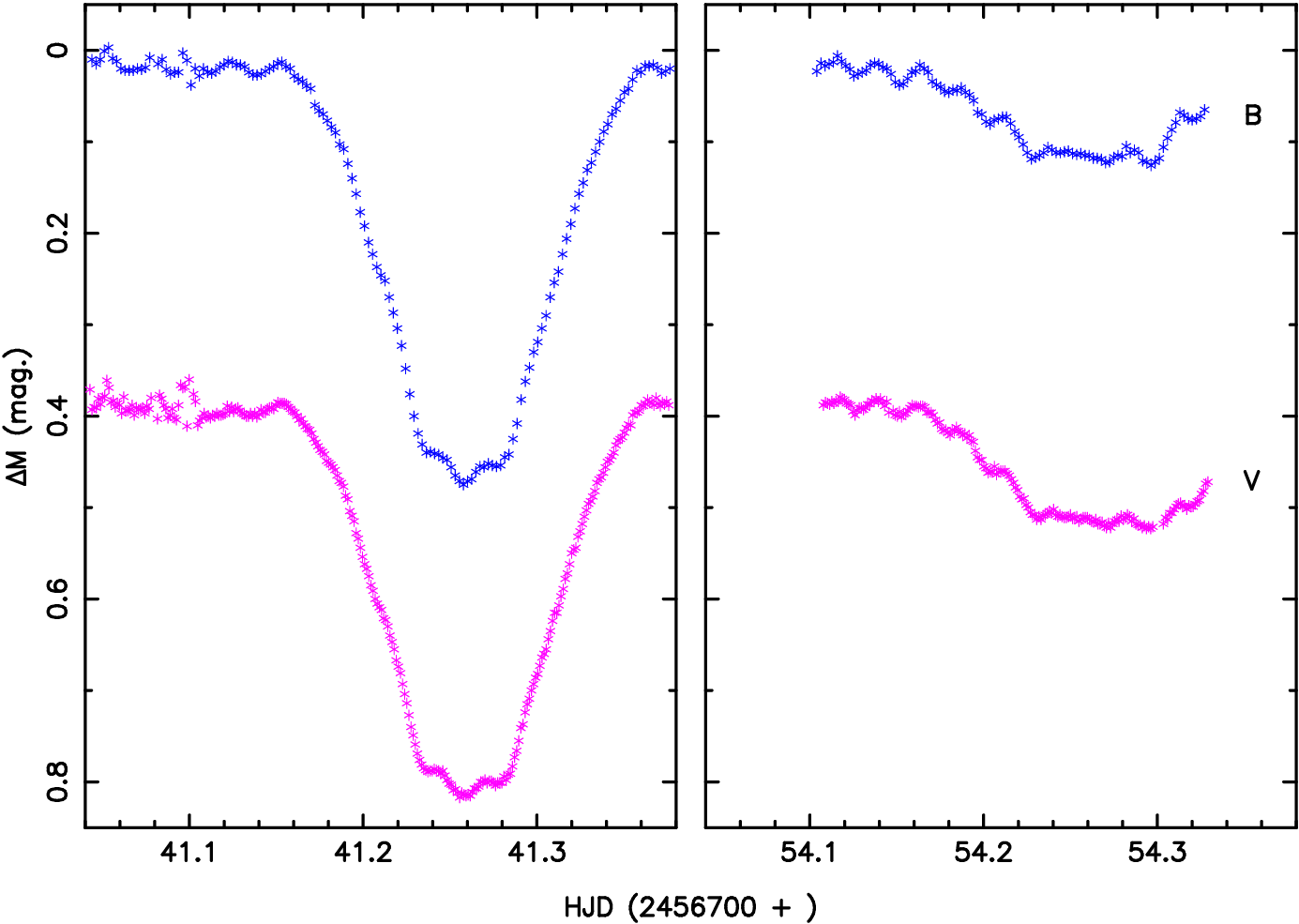}
\includegraphics[angle=270,scale=.6]{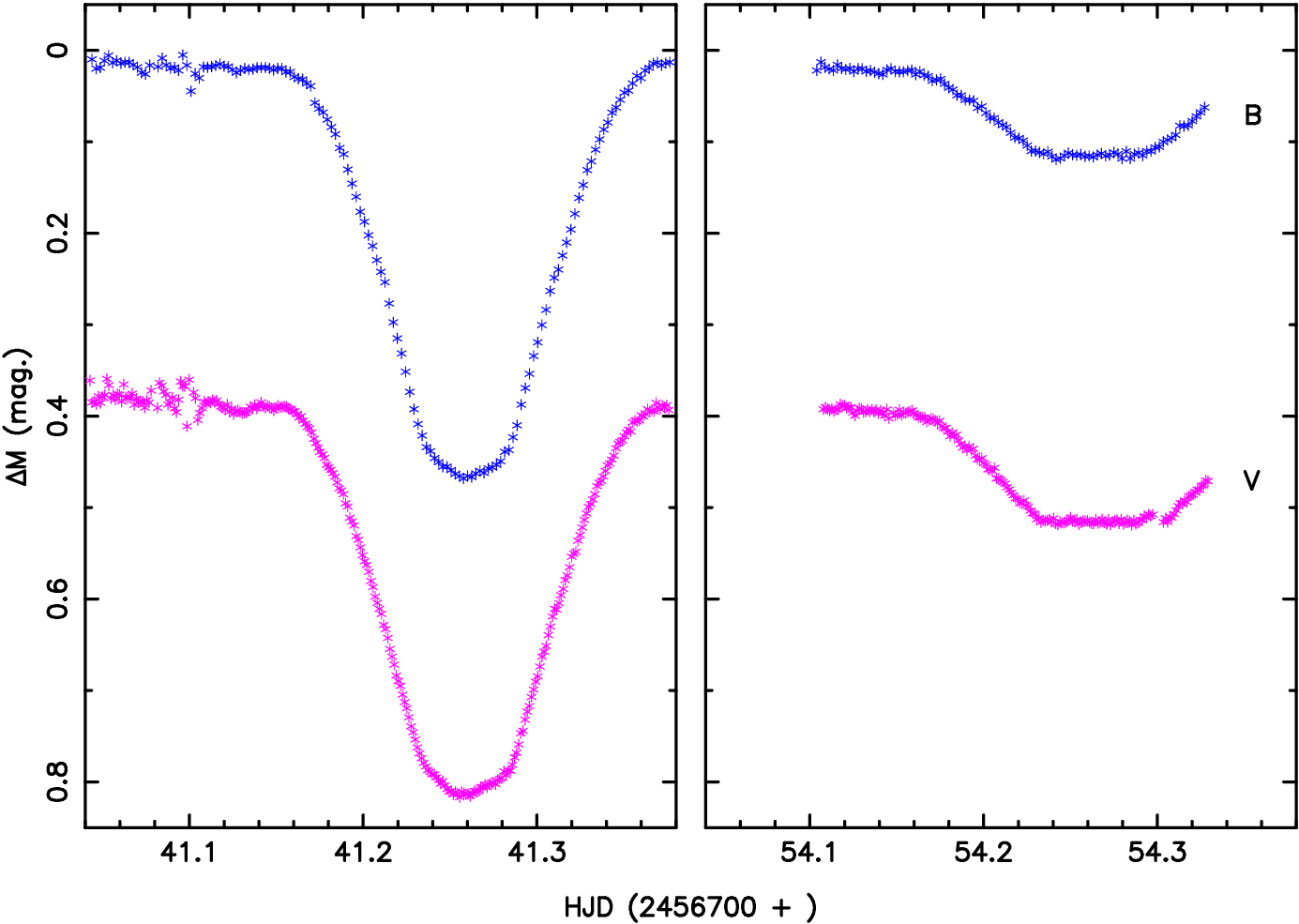}
\caption{The upper panels represent a segment of the original real-time B- and V-band light curves of OO Dra observed on March 24 and April 6, 2014. The lowers are those with short-term pulsations subtracted.}
\end{figure}

\section{The binary system}

A total of 3 completely covered eclipses, including 2 primary and 1 secondary ones were recorded in our data. By   using a least-squares parabolic fitting method, the epochs of these light minima were determined as given in Table 2. From the literatures (Lampens et al. 2010, Diethelm 2012, H\v{u}bscher et al. 2013), we collected 4 minima times for the star. Based on these data, the orbital period of the eclipsing binary was analyzed with the classical O-C method. It yields an orbital period of 1.238378 days for the binary system and the following linear ephemeris is given:
\begin{equation}
HJD(MinI)=2456741.2607(2)+1.238378(2)\times E
\end{equation}

\begin{table}
\begin{center}
\caption{Times of light minima of OO Dra and the residuals in respect to the derived ephemeris}
\begin{tabular}{lrrl}
\tableline\tableline
HJD & Epoch &  (O-C) & Ref\\
(HJD2450000+) & {} & {days} &{}\\
\tableline
4861.4032  &  -1518.0  &  -0.0001 & Lampens et al. 2010\\
5944.9791   &  -643.0   & -0.0047  & Diethelm 2012\\
6036.6300   &  -569.0    & 0.0063  & Diethelm 2012\\
6061.3904    & -549.0  &  -0.0009  & H\v{u}bscher \& Lehmann 2013\\
6741.2606    &    0.0  &  -0.0001   & Present study\\
6751.1675    &    8.0   & -0.0002   &Present study\\
6754.2633    &   10.5  &  -0.0004 &Present study\\
\tableline
\end{tabular}
\end{center}
\end{table}

By using the newly derived linear ephemeris, phases of all the measurements were computed. The folded light curves due to eclipsing were formed as shown in Figure 3. The general feature of the light curves are typical of Algol-type eclipsing binaries with nearly total eclipses, suggesting a large inclination close to 90$^{o}$ for the binary system. The depth of the primary eclipse was measured to be 0.46 mag. in B and 0.44 mag. in V. That of the secondary minima turned out to be 0.11 mag. in B, and 0.14 mag. in V, respectively.  

\begin{figure}
\centering
\includegraphics[angle=270,scale=.75]{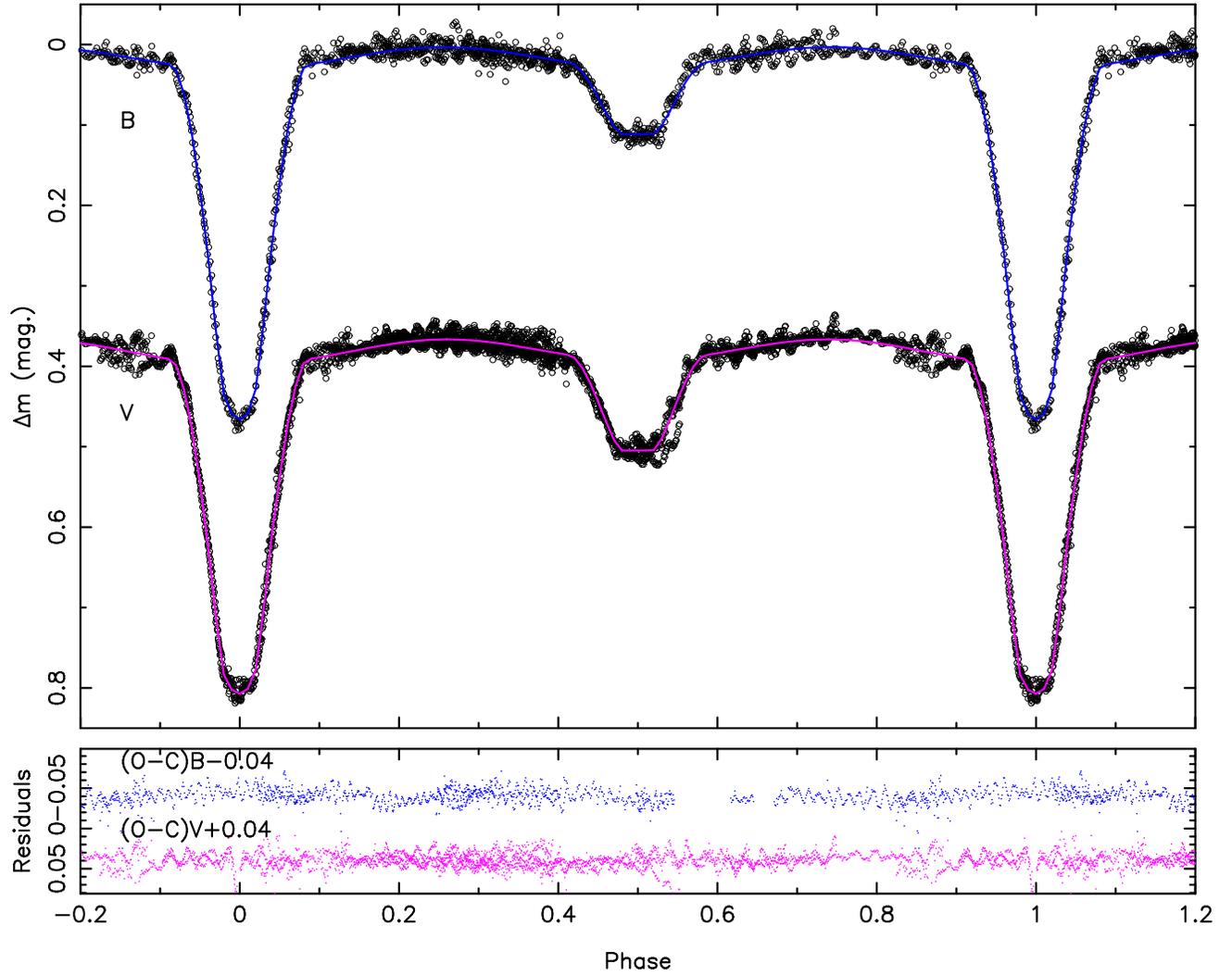}
\caption{Observed B- and V-band light curves of OO Dra and the theoretical synthesis}
\end{figure}

We made use of the Wilson-Devinney method (Wilson \& Devinney 1971, Wilson 1979) to obtain the first photometric solution for the binary system. The available B- and V-band light curves were simultaneously synthesized by applying the 2003 version of the WD code with the Kurucz atmospheres (Wilson 1990, Kallrath et al. 1998). A nonlinear limb-darkening law with the logarithmic form was applied in the light-curve synthesis. Considering of the short orbital period and the evolutionary status of classical Algols such as OO Dra, a circular orbit ($e=0$) and the synchronous rotation ($F_{2}=F_{2}=1.0$) assumption were accepted. The temperature of the massive (and usually luminous ) primary star was set at 8500 K adopted from Dimitrov et al. (2008). The initial bolometric ($X_{1}$, $X_{2}$, $Y_{1}$, $Y_{2}$) and monochromatic (x1, y1, x2, y2) limb-darkening coefficients of the components were taken from Van Hamme (1993). The gravity darkening exponents were set to be $g_{1}=1.0$ for the primary and $g_{2} = 0.32$ for the secondary component from Lucy (1967) according to their temperatures. The bolometric albedos were taken as $A_{1} = 1.0$ and $A_{2} = 0.5$ following Rucinski (1969).

The adjustable parameters in computing the photometric solutions are the orbital inclination (i), phase shift, mass ratio (q). surface temperature of the secondary ($T_{2}$), dimensionless luminosity of the primary ($L_{1}$) and the potentials of the two components ($\Omega_{1}$, $\Omega_{2}$). Since OO Dra is a single-line spectroscopic binary (Dimitrov et al. 2008), there is no mass ratio available from the radial-velocity solution. To search for an approximate mass ratio, we made a set of test solutions at the outset. The test solutions were computed at a series of assumed mass ratios with values ranging from 0.02 to 1.0.  The mass ratio was photometrically determined to be 0.097$\pm$0.001. The final best-fitting solution is given in Table 3. The synthesis of the observed B- and V-band light curves as well as the (O-C) residuals are shown in Figure 2. 

\begin{figure}
\includegraphics[angle=270,scale=.6]{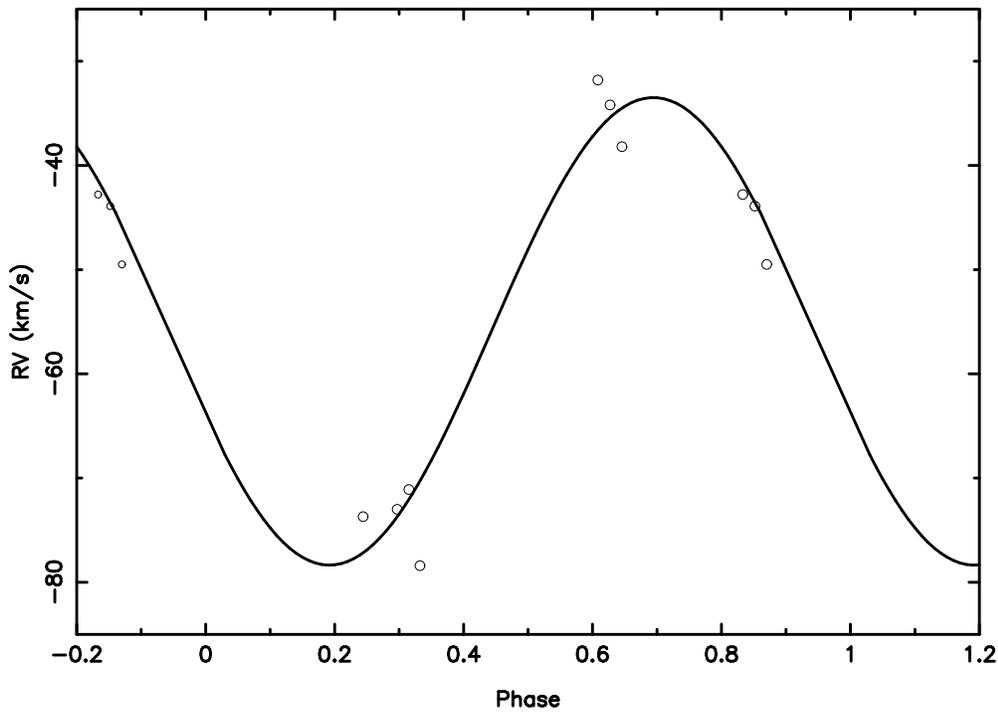}
\caption{Radial-velocity curve of OO Dra and the synthesis.}
\end{figure}

\begin{table}
\begin{center}
\caption{Light-curve and radial velocity solutions for the binary OO Dra}
\begin{small}
\begin{tabular}{lcc}
\tableline\tableline
Parameter & Best-Fit Value & Formal Error\\
\tableline
e* & 0 \\
$F_{1}=F_{2}$* & 1.0\\
$g_{1}$, $g_{2}$*& 1.0, 0.32\\
$A_{1}$, $A_{2}$*& 1.0, 0.50\\
$X_{1,bolo}$, $X_{2,bolo}$*& 0.658, 0.639\\
$Y_{1,bolo}$, $Y_{2,bolo}$*& 0.139, 0.241\\
$x_{1,B}$, $x_{2.B}$*& 0.785, 0.806\\
$y_{1,B}$, $y_{2.B}$*& 0.301, 0.232\\
$x_{1,V}$, $x_{2.V}$*& 0.676, 0.710\\
$y_{1,V}$, $y_{2.V}$*& 0.276, 0.275\\
$T_{1}$ (K)*  & 8500 & {}\\
i (deg)              & 85.66 & $\pm$0.14\\
$T_{2}$ (K)   & 6452 & $\pm$8\\
$\Omega_{1}$ & 3.211 & $\pm$0.007\\
$\Omega_{2}$ & 1.986 & $\pm$0.002\\
q=$M_{2}/M_{1}$ & 0.097 & $\pm$0.002\\
$\frac{L_{1}}{(L_{1}+L_{2})}_{B}$ & 0.922 & $\pm$0.001\\
$\frac{L_{1}}{(L_{1}+L_{2})}_{V}$ & 0.899 & $\pm$0.001\\
$P_{orb}$ (days) & 1.238378 & 0.000002\\
a ($R_{\odot}$) & 6.27 & $\pm$0.35\\
$\gamma$ (km/s) & -55.92 & $\pm$ 0.02\\
$M_{1}$ ($M_{\odot}$) & 1.97 & $\pm$0.25\\
$M_{2}$ ($M_{\odot}$) & 0.19 & $\pm$0.03\\
$R_{1}$ ($R_{\odot}$) & 2.04 & $\pm$0.09\\
$R_{2}$ ($R_{\odot}$) & 1.17 & $\pm$0.05\\
$log(L_{1}/L_{\odot})$ & 1.27 & $\pm$0.03\\
$log(L_{2}/L_{\odot})$ & 0.31 & $\pm$0.01\\
\tableline
\end{tabular}
\end{small}
\end{center}
*: assumed
\end{table}

The photometric solution reveals a detached configuration with the secondary nearly filling in its Roche lobe. The filling factor, defined as fraction of the stellar radius to that of the critical Roche lobe,  of the two components turned out to be 55.8\% and 97.5\%. This suggests that the binary could also be in semi-detached. To check this possibility, we have modeled the light curves with Mode 5 (semi-detached model with the star 2 filling its Roche lobe). It yields a converged solution at q=0.075, which also gives a satisfied fitting to the observations. But the sum of residuals, $\Sigma(O-C)^{2}$, is about 10\% larger than that of the solution with Mode 2 (detached model). We therefore adopted the results with detached configuration.

Based on the photometric solution, we have carried out a spectroscopic solution for OO Dra with the data published by Dimitrov et al. (2008). As it was indicated by the photometric solution that the less-massive secondary star contributes about 10 percent luminosity of the binary system, it is hard to be seen in the spectra.  
Radial velocities for only the primary component could be measured. By using the newly derived ephemeris, the radial velocity curve of the massive primary component was reformed. Two data points  detected at HJD2454600.4282 and HJD2454600.4516 were not  included into the analysis for the large scatterings. As a result, the radial-velocity curve modeling gives a  separation between the two components of $a=6.27\pm0.35R_{\odot}$ with a systematic velocity of  -55.92 km/s for the binary. With this, the absolute parameters of the components were then calculated as given in Table 3. The synthesis of the radial-velocity curve is shown in Figure 3. 

With the absolute parameters derived for OO Dra, the ratio of orbital  angular momentum to rotational angular momentum at equilibrium could be estimated for the binary system following Hut (1981):
\begin{equation}
\alpha=\frac{q}{1+q}\frac{1}{r_{g}^{2}}(\frac{a}{R_{1}})^{2}
\end{equation}
where $r_{g}$ is the gyration constant with a typical value of $r_{g}^{2}=0.1$ for main-sequence stars and varies negligibly during evolution. Taking the related parameters into Eq. (2), it gives a result of $\alpha \approx8.3$. It suggests that the rotational angular momentum is appreciable at equilibrium (Hut 1981) . This in turn supports the 
circular and synchronous assumptions we accepted during the solutions.

\section{The intrinsic pulsations}

In Figures 1 and 2, the pulsational light variability can be clearly seen in the phases of out-eclipses and the secondary minimum of the light curves. With the derived photometric solution, time-resolved theoretical light curves due to eclipsing were computed. Subtracting the eclipsing light changes from the original observational data, we detected the pure pulsational light variations of the primary star. This enables us to make a detailed analysis of the pulsation nature of OO Dra.

The frequency analysis was carried out with the algorithm Period04 (Lenz \& Breger 2005) based on the Fourier transform method. While doing that, we selected only those peaks with the signal to noise ratio (S/N) larger than 4.0 and appeared in both the B- and V-band data for further discussion. The noise levels were computed based on the residuals from the original data when all the trial frequencies were pre-whitened.  Figure 4 illustrates the Fourier analysis of both the B- and V-band data. The spectral window and the step-by-step amplitude spectra were plotted. Each spectrum panel in the figure corresponds to the residuals with all the previous frequencies pre-whitened. The bottom panels show the final residuals of the B- and V-band data, wherein the dashed lines represent the confidence curves with 4$\sigma$. The detected frequencies are listed in Table 4. 

\begin{figure}
\centering
\includegraphics[angle=270,scale=.75]{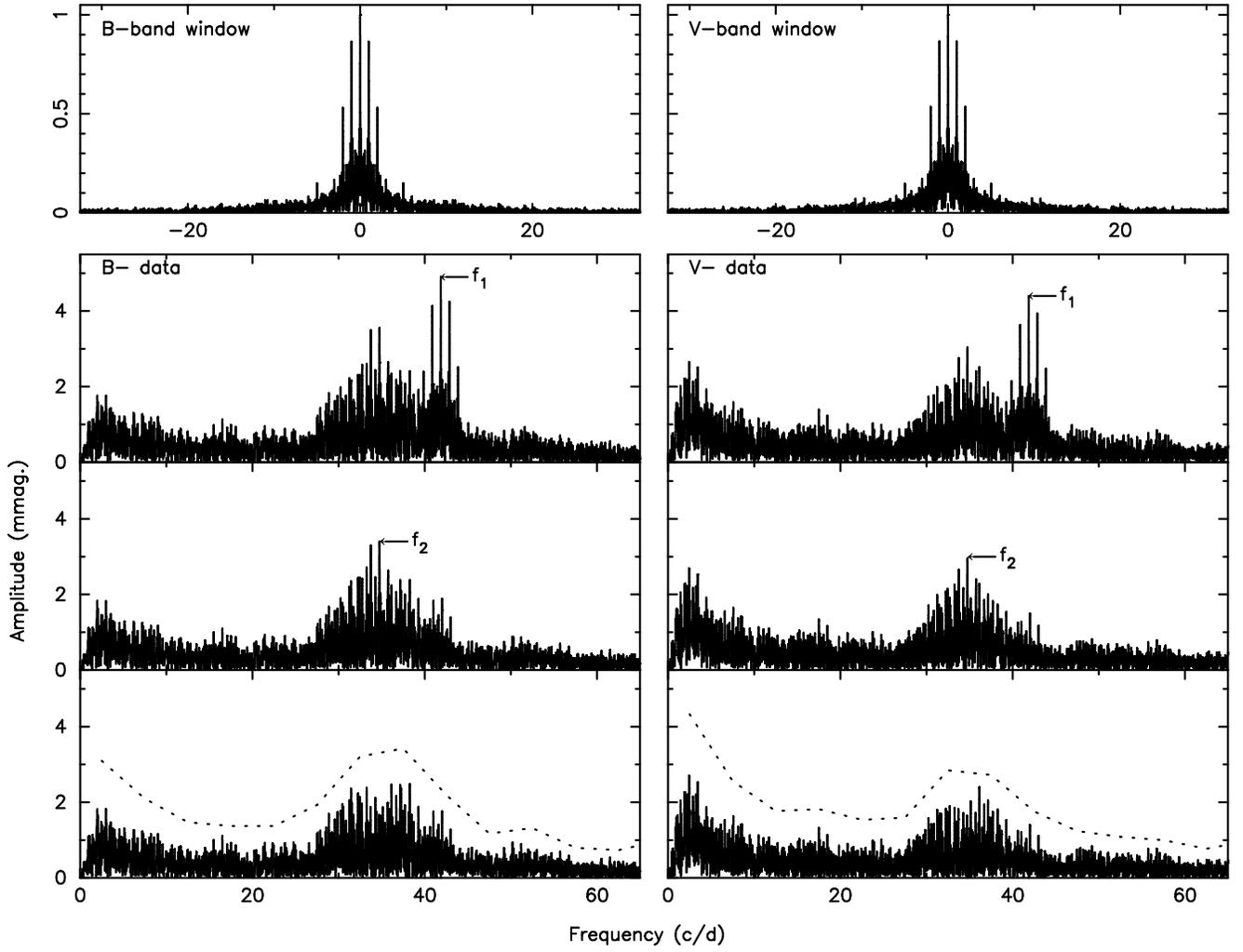}
\caption{Spectral windows and the step-by-step amplitude spectra of the intrinsic oscillations of OO Dra..}
\end{figure}

\begin{table}
\begin{center}
\caption{Pulsational properties of OO Dra}
\begin{tabular}{lcccc}
\tableline\tableline
Filter & Frequency & Amplitude & Phase & S/N\\
{} & (c/d) & (mmag) & (rad)\\
\tableline
B & 41.867$\pm$0.001 & 4.9$\pm$0.3  & 0.998$\pm$0.008 & 8.1\\
{} & 34.749$\pm$0.002 & 3.4$\pm$0.3  & 0.577$\pm$0.012 & 4.7\\
V & 41.865$\pm$0.001 & 4.2$\pm$0.2  & 0.063$\pm$0.007 & 8.2\\
{} & 34.753$\pm$0.002 & 2.8$\pm$0.2  & 0.185$\pm$0.010 & 4.3\\
\tableline
\end{tabular}
\end{center}
\end{table}

The periodograms show typical a feature of $\delta$ Sct stars with multi-periodicity. Two peaks at  $f_{1}$ = 41.87 c/d and $f_{2}$ = 34.75 c/d are clearly shown in both the B- and V-band original amplitude spectra. The dominant peak is 41.87 c/d, corresponding to a period of 0.02388 days, has a whole amplitude of about 9.8 mmag in B and 8.4 mmag in V. The amplitude of the second peak is measured to be 6.8 mmag in B and 5.6 mmag in V.  When the two frequencies were pre-whitened, a third frequency around 37 c/d arose. It is likely that one previously detected by Dimitrov et al. (2008). But it was not above the significance limit due to the high noise. This frequency is therefore not included into Table 4 for further discussion. In addition, there is no harmonics related to the rotation period detected, implying that the eclipses and proximity effects have been successfully removed from the data.

To check the result and discuss the pulsation nature further, we have computed the theoretical pulsation light curves using the adopted frequencies given in Table 4. After subtracted the pulsational light variations from the original observations, we detected the 'pure' light changes due to eclipsing. A segment of the eclipsing light curves were shown in the lower panels in Figure 1. It can be seen that the short-term light variations have been fairly removed. The totally eclipsing feature is clearly shown. A comparison between the two sets of light curves suggests that the intrinsic pulsations could be very probably from the hotter primary component of the system. This is identical with most eclipsing binaries with $\delta$ Sct variables. A discussion on the evolutionary status of the components in the next section also supports this possibility.

Because of the large uncertainty,  the absolute parameters derived for the components were not suitable for detailed analysis of the pulsation nature. Following the method described by Zhang et al. (2013), we used only the results from the photometric solution to calculate the mean density of pulsating primary star. The result is   $\rho_{1}/\rho_{\odot}=0.232\pm0.003$. Taking this value into the well-known equation, $Q=P_{pul}(\rho/\rho_{\odot})^{1/2}$, the pulsation constants of about 0.0116$\pm$0.0001 and 0.0139$\pm$0.0001 days were calculated for 41.87 c/d and 34.75 c/d oscillations, respectively. By using the FAMIAS (Frequency Analysis and Mode Identification for Asteroseismology) program (Zima, 2008), we have tried to make a preliminary mode identification. It suggests an angular quantum number of $l=0$ or $l=1$, implying that the star could be  pulsating in radial mode or low degree non-radial modes. If this is the case, the detected frequencies could be further identified as the 5th and/or 4th overtone modes (Fitch 1981).

\section{Summary and discussion}

We have presented the simultaneous B- and V-band photometry of the Algol-type eclipsing binary OO Dra. It confirms  $\delta$ Sct-type pulsation nature of the eclipsing binary. The first photometric solution and physical parameters of the binary system were determined.  The intrinsic pulsational characteristics of the primary component were analyzed.

The light curves of OO Dra show totally eclipses. This enables us to obtain a reliable photometric solution for the binary system through light modeling by using the Wilson-Devinney method to determine the geometric parameters. The photometric solution reveals a very small mass ratio of 0.097. This could be the least mass ratio after KIC10661783 (Southworth et al. 2011) among the known eclipsing binaries containing of $\delta$ Sct-type pulsating components. The light-curve synthesis indicates a detached configuration for the binary system though the secondary component is very nearly filling its Roche lobe. It means that OO Dra could not be a member of the oEA (Mkrtichian et al. 2004) class. Since the number of $\delta$ Sct stars in confirmed detached Algols is very small (Liakos et al. 2012), this star could be an important object in studying of pulsating eclipsing binaries.

With results from the photometric solution, a radial-velocity synthesis was further performed with the measurements published by Dimitrov et al. (2008), and the absolute parameters of the components of the binary were calculated. The results show that the primary component is almost an un-evolved main-sequence star, the secondary is while quite evolved in over-luminous and over-sized. Comparison with the mass-luminosity relations of Algols (\.{I}bano\v{g}lu et al. 2006), we find that the mass and radius detected for the primary component of OO Dra match well with that of the detached Algols ($L_{1}\propto M_{1}^{3.92\pm0.05}$) rather that the semi-detached ones ($L_{1}\propto M_{1}^{3.20\pm0.25}$). This in turn supports the detached configuration as indicated by the photometric solution.

Based on the photometric solution, we have calculated the theoretical light curves due to eclipsing. Subtracting the theoretical eclipsing light changes from the observations, the intrinsic pulsational light variations from the hotter primary component were picked up. A frequency analysis of the remaining residuals showed multiple periodicities of the pulsating star. Two confident frequencies at 41.87 and 34.75 c/d were detected in both filters. The dominant pulsation period was determined to be 0.02388 days. The ratio of the pulsational to orbital period of the star turned out to be $P_{pul}/P_{orb}=0.0193$. It agrees well with the correlation between the orbital and pulsational periods preposed by Zhang et al. (2013) for eclipsing binaries containing of $\delta$ Sct-type components.

With the mean density of the primary component deduced from photometric solution, the pulsation constants of the detected  frequencies were computed and a preliminary mode identification was made. The pulsation constant of the dominant frequency was computed to be about 0.012 days, which is very close to the mean value of $\delta$ Sct stars in eclipsing binaries (Zhang et al. 2013). It also suggests that the primary component of OO Dra could pulsate in fifth or fourth overtone assuming that it is in radial or low degree non-radial modes as most $\delta$ Sct stars do.

\acknowledgments

This work is supported by the National Natural Science Foundation of China (NSFC) and the NSFC/CAS Joint Fund of Astronomy through grants 11373037, U1231202 and 2013CB834900. LD would like to acknowledge partial support by National Key Basic Research Program of China 2014CB845703. The authors are grateful to the anonymous referee for the valuable comments.

\end{document}